\documentclass{article}
\usepackage{ijcai25}

\usepackage{times}
\usepackage{soul}
\usepackage{url}
\usepackage[hidelinks]{hyperref}
\usepackage[utf8]{inputenc}
\usepackage[small]{caption}
\usepackage{graphicx}
\usepackage{amsmath}
\usepackage{amsthm}
\usepackage{booktabs}
\usepackage{algorithm}
\usepackage{algorithmic}
\usepackage[switch]{lineno}

\usepackage[x11names]{xcolor}
\usepackage{amssymb}
\usepackage{pifont}
\usepackage{amsmath}
\usepackage{multirow} 
\usepackage{multicol} 
\usepackage{makecell} 
\usepackage{booktabs}
\usepackage{threeparttable}
\usepackage{xcolor}
\definecolor{colornode}{RGB}{30, 50, 100} 
\definecolor{coloredge}{RGB}{80, 120, 20} 
\definecolor{colorglobal}{RGB}{30, 30, 30} 

\title{PerfSeer: An Efficient and Accurate Deep Learning Models Performance Predictor}

\author{
Xinlong Zhao$^1$\and
Jiande Sun$^1$\and
Jia Zhang$^1$\and
Sujuan Hou$^1$\and
Shuai Li$^2$\and
Tong Liu$^3$\And
Ke Liu$^1$\thanks{Corresponding author.}
\affiliations
$^1$Shandong Normal University\\
$^2$Shandong University\\
$^3$IEIT SYSTEMS Co., Ltd.\\
\emails
xinloongzhao@gmail.com,
jiandesun@hotmail.com,
\{zhangjia, sujuanhou\}@sdnu.edu.cn,
shuaili@sdu.edu.cn,
tong.liu@inspur.com,
sdu\_liuke@outlook.com
}

\begin{document}

\maketitle

\begin{abstract}
Predicting the performance of deep learning (DL) models, such as execution time and resource utilization, is crucial for Neural Architecture Search (NAS), DL cluster schedulers, and other technologies that advance deep learning.
The representation of a model is the foundation for its performance prediction. However, existing methods cannot comprehensively represent diverse model configurations, resulting in unsatisfactory accuracy.
To address this, we represent a model as a graph that includes the topology, along with the node, edge, and global features, all of which are crucial for effectively capturing the performance of the model.
Based on this representation, we propose PerfSeer, a novel predictor that uses a Graph Neural Network (GNN)-based performance prediction model, SeerNet. 
SeerNet fully leverages the topology and various features, while incorporating optimizations such as Synergistic Max-Mean aggregation (SynMM) and Global-Node Perspective Boost (GNPB) to capture the critical performance information more effectively, enabling it to predict the performance of models accurately.
Furthermore, SeerNet can be extended to SeerNet-Multi by using Project Conflicting Gradients (PCGrad), enabling efficient simultaneous prediction of multiple performance metrics without significantly affecting accuracy.
We constructed a dataset containing performance metrics for 53k+ model configurations, including execution time, memory usage, and Streaming Multiprocessor (SM) utilization during both training and inference.
The evaluation results show that PerfSeer outperforms nn-Meter, Brp-NAS, and DIPPM.
\end{abstract}

\section{INTRODUCTION}
Deep learning (DL) has achieved significant success across various fields \cite{RNN,cnn}. Understanding the performance of models (e.g., execution time and resource utilization) is crucial for technologies such as Neural Architecture Search (NAS) and DL cluster schedulers, which drive the advancement of deep learning.
NAS relies on performance metrics like execution time \cite{Brp-nas,latencyaware-nas} and memory usage \cite{sas-nas,micronets-nas} to design efficient models that meet hardware constraints. Similarly, efficient DL cluster schedulers use these metrics to reduce job completion time and improve accelerator utilization \cite{Liquid,Horus}.

Acquiring performance metrics through online profiling is both time-consuming and costly. Consequently, various predictors have been proposed to predict the model performance, generally categorized into three main methods.
The first method \cite{Liquid,Horus} treats the model as a whole, representing it with global features like floating point operations (FLOPs). These features are then used in a basic model, such as Multi-Layer Perceptron (MLP) \cite{mlp}, to predict the performance.
The second method \cite{justus,Nn-meter} divides the model into multiple parts (e.g., operations or units), each represented by specific features. These features are then fed into a similar basic model to predict the performance of each part, which is aggregated to estimate the overall performance.
The third method \cite{Brp-nas,dnnperf,DIPPM,PerfSAGE} treats the model as a computational graph, represented by its topology and node features. These features are then input into a GNN-based model. This approach enables more accurate predictions by learning the performance of each operation and their interdependencies.
Overall, 
performance predictors use machine-readable data to represent models and then feed this data into prediction models to predict the performance.

However, the model representation in these predictors is not comprehensive enough, resulting in the loss of important information related to model performance, leading to suboptimal prediction accuracy.
Specifically, the node, edge, and global features are all crucial for a model. Node features capture the characteristics of each operation. Edge features describe the data flow and dependencies between operations. Global features provide a view of the overall structure and complexity of the model.
Existing predictors have not selected and utilized these features effectively. 

We propose PerfSeer, a novel predictor that can comprehensively represent the model and fully leverage this representation to accurately predict the performance of models.

Initially, 
PerfSeer uses Graph Extractor to parse the computational graph of a model and generate a performance graph (PerfGraph), which can represent the model performance.
This graph, in addition to the topology, also contains the node, edge, and global features, all of which improve the performance prediction accuracy.
Notably, we construct several unique and effective features, such as arithmetic intensity and statistics of computation and memory access, which are not used by any previous predictors and have been empirically proven to enhance prediction accuracy noticeably.

Subsequently, 
the PerfGraph is fed into our proposed prediction model, SeerNet, to predict the performance.
SeerNet, inspired by \cite{gn}, enables each feature to update its own representation by utilizing both its own features and aggregated information from other features.  
This allows SeerNet to not only learn the topology but also fully leverage various features, providing a comprehensive understanding of the model. 
Specifically, it learns computation and memory access from the node features, data flow and dependencies between operations from the edge features, and the structure and complexity of the entire model from the global features.
To capture the performance information of models more effectively, SeerNet employs Synergistic Max-Mean aggregation (SynMM) and Global-Node Perspective Boost (GNPB).
SynMM can better aggregate the node features, generating a more comprehensive and robust model representation.
GNPB enables complementary learning between the node and global features, mutually enriching their perspectives.

In addition, 
We propose SeerNet-Multi, a multi-metric performance prediction model designed for applications that require simultaneous consideration of multiple metrics (e.g., \cite{Horus,Liquid}). Training and maintaining separate models for each metric is inefficient, so SeerNet-Multi extends SeerNet with multiple prediction heads to predict multiple metrics simultaneously. To address conflicts in parameter updates caused by differing gradient directions across tasks, we integrate Projecting Conflicting Gradients (PCGrad) \cite{pcgrad}, ensuring high accuracy. 
As a result, SeerNet-Multi provides efficient and accurate multi-metric predictions.

We constructed a dataset with over 53k model configurations, covering key performance metrics such as execution time, memory usage, and Streaming Multiprocessor (SM) utilization during both training and inference in Nvidia GeForce RTX 3090. 
The dataset spans various architectures, such as VGG \cite{vggnet}, GoogLeNet \cite{googlenet}, ResNe(X)t \cite{resnet,resnext}, MobileNet \cite{Mobilenets}, and DenseNet \cite{densenet}. It also includes a wide range of FLOPs, from 49M to 22T.
Evaluation results show that SeerNet achieves an average Mean Absolute Percentage Error (MAPE) of 5.14\%, while SeerNet-Multi achieves 7.75\% MAPE, outperforming other predictors.

We summarize our key contributions as follows:
\begin{itemize}
    \item We propose a novel performance predictor, PerfSeer.
    Experiments show that PerfSeer, with low deployment and usage overhead, predicts multiple performance metrics accurately for models of various architectures and frameworks on different devices during training and inference, enabling broad applicability.

    \item We design the performance prediction models, SeerNet and SeerNet-Multi.
    SeerNet achieves state-of-the-art prediction accuracy. 
    It rigorously selects, novelly constructs, and fully leverages the performance-related node, edge, and global features. Additionally, it enhances performance capture through SynMM and GNPB.
    SeerNet-Multi efficiently predicts multiple performance metrics without significantly affecting accuracy by addressing conflicting gradient issues across tasks using PCGrad.  

    \item We construct a performance dataset\footnote [1]{https://github.com/upuuuuuu/PerfSeer}.
    The dataset, which contains over 53k model configurations, covers key performance metrics across various architectures and FLOPs during both training and inference.
\end{itemize}

\begin{figure*}[!htb]
  \centering
  \includegraphics[width=2.0\columnwidth]{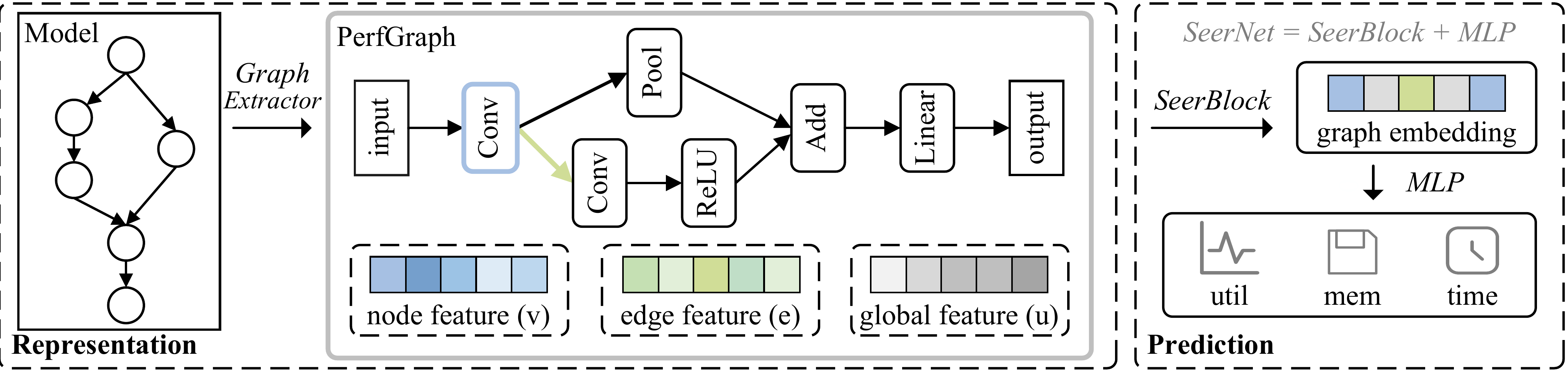}
  \caption{Overview of PerfSeer. 
    Node features are shown in \textcolor[RGB]{166, 192, 227}{"blue"}, edge features in \textcolor[RGB]{208, 221, 151}{"green"}, and global features in \textcolor[RGB]{120, 120, 120}{"gray"}.}
  \label{fig:overview}
\end{figure*}

\section{Related Work} \label{sec:related work} 
\paragraph{Global Features-Based Predictor.} 
Liquid \cite{Liquid} and Horus \cite{Horus} use traditional models, such as Random Forest (RF) \cite{randomforest} and XGBoost \cite{Xgboost}, to predict GPU utilization and memory usage during model training by inputting global features like batch size, FLOPs, and parameter size. However, these methods are limited as they fail to capture internal execution details, and different models can exhibit vastly different performance, even with the same global features.

\paragraph{Operations/Unit-Wise Predictor.}
Justus et al. \cite{justus} predict the execution time of training models on edge devices by predicting the execution time of each operation and aggregating them. They use hardware, operations, and specific features as inputs to an MLP. nn-Meter \cite{Nn-meter} splits the model inference into kernels, predicting the execution time of each by feeding features like hyper-parameters, FLOPs, and input/output shapes into an RF model. This execution time is aggregated to get the total execution time. However, this approach relies on heuristic detection functions that require a deep understanding of execution details across different models and devices, which can be costly and inaccurate, leading to inaccurate predictions.

\paragraph{GNN-Based Predictor.}
BRP-NAS (Eagle) \cite{Brp-nas} uses a Graph Convolutional Network (GCN) \cite{GCN}-based model to predict execution time and accuracy for inference models, applied to NAS on NAS-Bench-201 \cite{nas-bench}. 
DNNPerf \cite{dnnperf} uses node and edge features in a custom Graph Attention Network (GAT) \cite{GAT}-based model to predict execution time and memory usage for training models. 
DIPPM \cite{DIPPM} feeds node and global features into a Graph Sample and Aggregation (GraphSAGE) \cite{GraphSage}-based model (PMGNS), which generates embeddings and uses multiple prediction heads to predict execution time, memory usage, and energy consumption for inference models. 
This approach captures execution details by learning both the topology and node features of models, ensuring prediction accuracy.

\section{PerfSeer DESIGN}
PerfSeer consists of two main parts: representation and prediction, as shown in Figure \ref{fig:overview}.
\begin{enumerate}
\item
\textbf{Representation.} PerfSeer uses Graph Extractor to analyze the computational graph of a model, generating a performance graph (PerfGraph) to represent the model. The PerfGraph encompasses the topology as well as the node, edge, and global features, comprehensively preserving the performance information of models 
(Section \ref{sec:graph extractor}).

\item
\textbf{Prediction.} PerfSeer employs SeerNet and SeerNet-Multi, our designed prediction models. These models effectively leverage the extracted topology and various features. They capture the critical performance information to predict metrics like execution time, memory usage, and SM utilization (Section \ref{sec:seernet}).
\end{enumerate}

Through these two parts, PerfSeer can comprehensively represent the model and fully leverage this representation to predict the model performance accurately.

\subsection{Representation}\label{sec:graph extractor}
We use the Graph Extractor via ONNX to generate a PerfGraph, extracting the topology and as many performance-related features as possible to preserve and represent the performance information of models. 
PerfSeer is compatible with multiple DL frameworks, such as PyTorch, TensorFlow, and MXNet, unlike other predictors that support only a few.
\subsubsection{Definition of PerfGraph.}\label{sec:definition of graph}
PerfGraph is defined as a 3-tuple $G = \left(\uu, V, E\right)$.
The $\uu$ represents the global features, which are the features of the entire model.
The $V = \left\{\vv_i\right\}_{i=1:N^v}$ is the set of nodes ($N^v$ is the number of nodes), where $\vv_i$ represents the features of node $i$ (Node $i$ represents operation node $i$ in the computational graph).
The $E = \left\{\left(\ee_j, s_j, t_j\right)\right\}_{j=1:N^e}$ is the set of edges ($N^e$ is the number of edges), where $s_j$ is the index of the source node and $t_j$ is the index of the target node, indicating a directed edge $j$ from the source node $s_j$ to the target node $t_j$. 
Edge $j$ represents the data flow from operation $s_j$ to operation $t_j$, where operation $t_j$ depends on operation $s_j$, and \(\ee_j\) represents the features of edge $j$.

\subsubsection{Features of PerfGraph.} \label{sec:features of graph}
The features of the PerfGraph consist of the node, edge, and global features, described as follows:

\begin{figure*}[!htb]
  \centering
  \includegraphics[width=2.05\columnwidth]{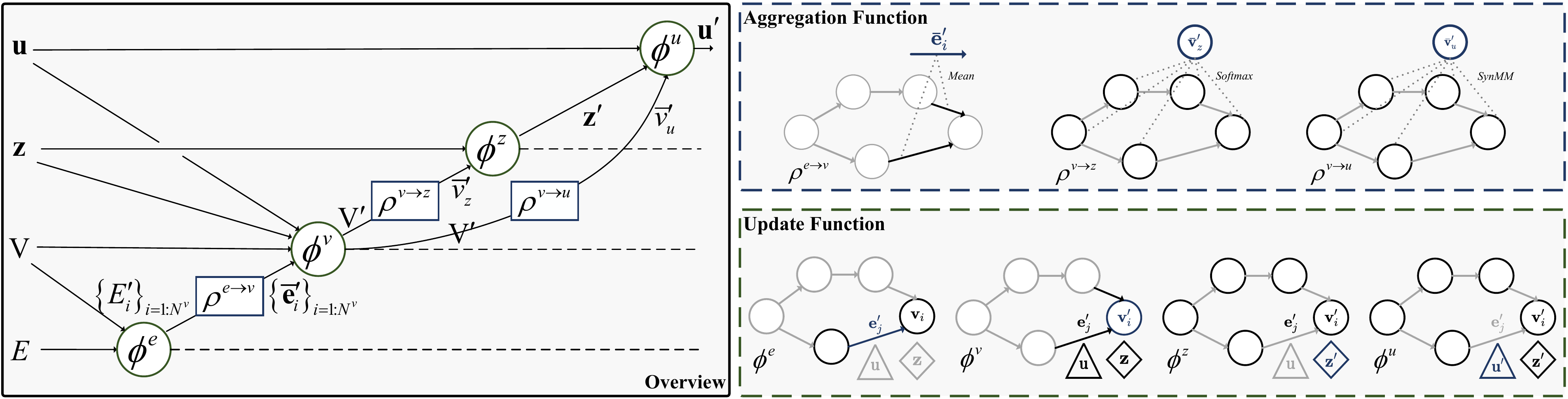}
  \caption{Architecture of SeerBlock. The target features for aggregation or update are represented in \textcolor[RGB]{70, 114, 196}{"blue"}; the features required during the aggregation or update are represented in "black"; irrelevant features are represented in \textcolor[RGB]{120, 120, 120}{"gray"}. 
  The function $\phi^x$ represents the update function for feature $x$, while $\rho^{x \rightarrow y}$ represents the aggregation function for feature $x$, with the aggregated result used to update feature $y$.}
  \label{fig:seerblock}
\end{figure*}

\emph{{Node Features.}}
Node features are represented as $\vv = \left(\vv^{hp} \cat \vv^{c} \cat \vv^{m} \cat \vv^{a} \cat \vv^{p} \right)$. $\cat$ represents the concatenation, responsible for concatenating different features.

$\vv^{hp}$ represents the hyper-parameters of the node, such as the kernel size of the convolution, and serves as the fundamental representation of the node, with all other features derived from it.

$\vv^{c}$ denotes the computation information (i.e., FLOPs) of the node, representing the computational requirements of the node.

$\vv^m$ represents the memory access information of the node, including the memory access cost (MAC) and the weight size.
The MAC is the sum of the sizes of the input, weight, and output tensors.
The features symbolize the memory access requirements of the node.

$\vv^{a}$ represents arithmetic intensity, defined as the ratio of FLOPs to MAC. The features can distinguish whether the node is arithmetic-intensive or memory-intensive.

$\vv^{p}$ denotes the proportions of FLOPs, MAC, and weight size relative to the model. The features can indicate the contribution of the node to the computational and memory requirements of the model.

\emph{{Edge Features.}}
Edge features are represented as $\ee = \left( \ee^{sz} \cat \ee^{sp} \right)$. 

$\ee^{sz}$ represents the size of the tensor delivered by the edge. 

$\ee^{sp}$ represents the shape of the tensor delivered by the edge, which includes the batch size, channel, height, and width. 

\emph{{Global Features.}}
Global features are represented as $\uu = \left( \uu^{gp} \cat \uu^{c} \cat \uu^{m} \cat \uu^{a} \cat \uu^{b} \right)$.

     $\uu^{gp}$ includes the number of nodes, edges, and density. The number of nodes ($N^v$) and edges ($N^e$) represent the counts of computations and memory accesses in the model, respectively. Density, defined as \(\frac{N^e}{N^e(N^e-1)}\), reflects the interconnectedness of the nodes in the graph, indicating how densely the nodes are connected. The features provide a topological profile of the entire computational graph of the model.

    $\uu^{c}$ represents FLOP statistics for all nodes in the model, including total, average, median, and maximum FLOPs. The features capture the computational characteristics of the model.
    
    $\uu^{m}$ represents memory access statistics for all nodes, such as total, average, median, and maximum values for MAC and weight size, as well as the average tensor size per edge, reflecting the memory access profile of the model.
    
    $\uu^{a}$ denotes arithmetic intensity, the ratio of FLOPs to MAC, which helps distinguish whether the model is arithmetic-intensive or memory-intensive.
    
    $\uu^{b}$ represents the batch size, which indicates the number of samples processed simultaneously by the model.

\subsection{Prediction}\label{sec:seernet}
SeerNet comprises a SeerBlock and an MLP-based prediction head.
The PerfGraph is input into SeerBlock, which learns a graph embedding that captures the performance information from the node, edge, global features, and the topology of the computational graph. This embedding is then passed to the prediction head, a two-layer MLP with 256 hidden channels, to predict the performance accurately.
\subsubsection{SeerBlock.}\label{sec:seerblock}
SeerBlock, inspired by \cite{gn}, enables each feature to update its own representation by utilizing both its own features and the aggregated information from other features.
The workflow of SeerBlock, shown in Figure \ref{fig:seerblock}, consists of four update functions ($\phi$) and three aggregation functions ($\rho$). 
\begin{align}
  \begin{split}
    \ee'_j &= \phi^e\left(\ee_j, \vv_{s_j}, \vv_{t_j}, \uu \right) \\
    \vv'_i &= \phi^v\left(\mathbf{\bar{e}}'_i, \vv_i, \zz, \uu \right) \\
    \zz' &= \phi^z\left(\mathbf{\bar{v}_z}', \zz \right) \\
    \uu' &= \phi^u\left(\mathbf{\bar{v}_u}', \zz, \uu \right) \\
  \end{split}
  \begin{split}
    \mathbf{\bar{e}}'_i &= \rho^{e \rightarrow v}\left(E'_i\right) \\
    \mathbf{\bar{v}}'_z &= \rho^{v \rightarrow z}\left(V'\right)  \\
    \mathbf{\bar{v}}'_u &= \rho^{v \rightarrow z}\left(V'\right)  \\
  \end{split}
  \label{eq:seerblock}
\end{align}

Where $E'_i = \left\{\left(\ee'_j, s_j, t_j \right)\right\}_{t_j=i,\; j=1:N^e}$, $V'=\left\{\vv'_i\right\}_{i=1:N^v}$, and $\zz$ represents the features of the global node (Section \ref{sec:gnpb}). The details are as follows:

\begin{enumerate}
\item $\phi^e$ is applied to per edge, to compute the updated edge features, $\ee'_j$.
The set of resulting per-edge outputs for each node, $i$, is,
$E'_i = \left\{\left(\ee'_j, s_j, t_j \right)\right\}_{t_j=i,\; j=1:N^e}$.
\begin{equation}
  \ee'_j = \phi^e\left(\ee_j, \vv_{s_j}, \vv_{t_j}\right) = \mathrm{MLP}_e\left(\ee_j \cat \vv_{s_j} \cat \vv_{t_j}\right).
\end{equation}

\item $\rho^{e\rightarrow v}$ is applied to $E'_i$, and aggregates the edge updates for the edges that project to node $i$, into $\mathbf{\bar{e}}'_i$, which will be used in node update in the next step. 
\begin{equation}
    \mathbf{\bar{\ee}'}_i = \rho^{e\rightarrow v}\left(E'_i\right) = \frac{1}{|E'_i|}\sum_{e'_j \in E'_i} \ee'_j.
\end{equation}

\item $\phi^v$ is applied to each node $i$,
to compute the updated node features, $\vv'_i$. 
The set of resulting per-node outputs is, $V'=\left\{\vv'_i\right\}_{i=1:N^v}$.
\begin{equation}
    \vv'_i = \phi^v\left(\mathbf{\bar{e}}'_i, \vv_i, \zz, \uu\right) = \mathrm{MLP}_v \left(\bar{\ee}'_i \cat (\vv_i + \zz) \cat \uu\right).
    \label{eq:node update}
\end{equation}

\item $\rho^{v \rightarrow z}$ is applied to $V'$, and aggregates all node updates, into $\bar{\vv}_z'$, which will then be used to update the global node in the next step (Section \ref{sec:gnpb}). 
\begin{equation}
    \bar{\vv}_z' = \rho^{v\rightarrow z}\left(V'\right) = {\text{softmax}}(V'). 
    \label{eq:aggr v2n}
\end{equation}

\item $\phi^z$ is applied once per graph, 
to compute the updated global node features, $z'$. 
\begin{equation}
    z' = \phi^z\left({\bar{\vv}}_z', \zz\right) = \mathrm{MLP}_z\left({\bar{\vv}}_z' + \zz\right).
    \label{eq:global node update}
\end{equation}

\item $\rho^{v \rightarrow u}$ is applied to $V'$, and aggregates all node updates, into $\bar{\vv}_u'$, which will then be used in the update of global features in the next step (Section \ref{sec:synmm}). 
\begin{equation}
    \bar{\vv}_u' = \rho^{v\rightarrow u}\left(V'\right) = \text{SynMM}\left(V'\right).
\end{equation}

\item $\phi^u$ is applied once per graph, 
to compute the updated global features, $\uu'$. $\uu'$ is the output of the SeerBlock.
\begin{equation}
    \uu' = \phi^u\left(\bar{\vv}_u', \mathbf{z'}, \uu\right) = \mathrm{MLP}_u\left(\bar{\vv}_u' \cat \mathbf{z'} \cat \uu\right).
\end{equation}

\end{enumerate}
Each MLP in SeerBlock has one layer with 256 channels, except for $\text{MLP}_z$, which has two layers. Given the PerfGraph $G = \left(\uu, V, E\right)$, SeerBlock produces the updated PerfGraph $G' = \left(\uu', V', E'\right)$, and the graph embedding $\uu'$ is fed into the prediction head.

\begin{figure}[!htb]
  \centering
  \includegraphics[width=1.0\columnwidth]{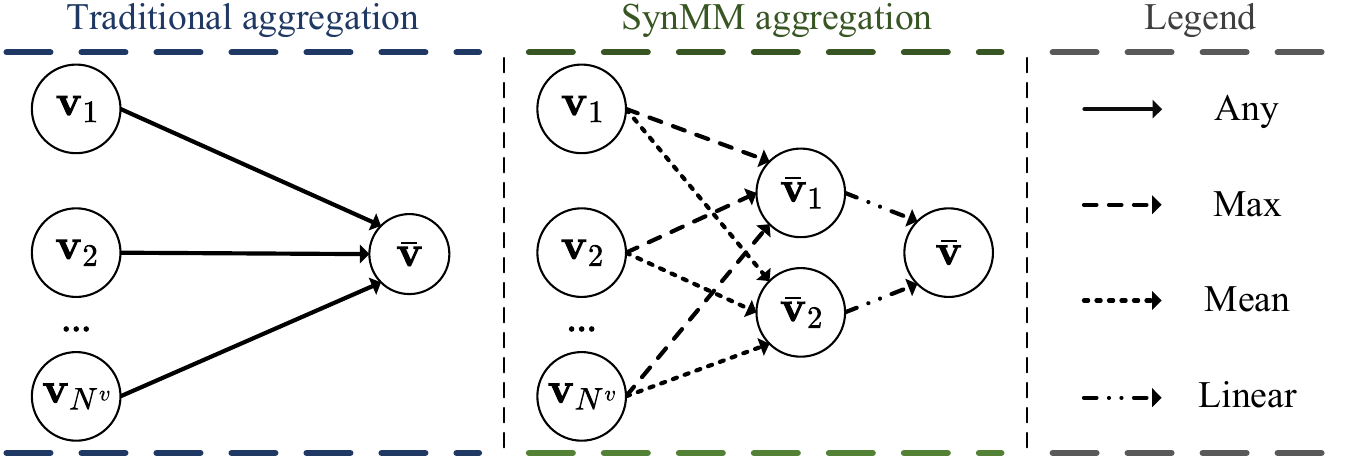}
  \caption{SynMM aggregation.}
  \label{fig:readout}
\end{figure}
\subsubsection{Synergistic Max-Mean Aggregation.}\label{sec:synmm}
SynMM is the customized function for the node features aggregation. As shown in Figure \ref{fig:readout}, the node features are first fed into the max and mean aggregation to obtain intermediate results, $\bar{\vv}_1$ and $\bar{\vv}_2$. 
Then, linear aggregation combines these results to generate the final aggregated features, $\bar{\vv}$. 
Max aggregation extracts the most prominent features from crucial nodes, while mean aggregation captures global information and overall model characteristics. 
Linear aggregation consolidates the results from both max and mean aggregation, providing a more comprehensive and robust model representation.

\subsubsection{Global-Node Perspective Boost.}\label{sec:gnpb}
GNPB, based on \cite{virtualnode}, not only introduces a global node connected to each node, with $\zz$ representing its features, but also designs an initialization method, updates its features, and incorporates them into the final graph embedding update.
The process is as follows:
First, the $\zz$ are initialized using ${\text{softmax}}$, aggregating the initial features of all nodes to capture global information.  
During the node update, each node utilizes the $\zz$ to incorporate global information (Equ. \ref{eq:node update}).  
The $\zz$ then updates itself by aggregating information from all updated nodes (Equ. \ref{eq:global node update}).
The global node offers a unique global perspective by aggregating information from all nodes, distinct from the global features based on the inherent characteristics of models. GNPB enables complementary learning between the node and global features, enriching both perspectives.

\subsubsection{SeerNet-Multi.}\label{sec:seernet multi}
SeerNet-Multi extends SeerNet by using multiple MLP-based prediction heads to predict different performance metrics separately.
In addition, PCGrad is employed to mitigate gradient conflicts among different performance metrics prediction tasks during the training of SeerNet-Multi. PCGrad works by adjusting the gradients of each task to minimize conflicts that arise when gradients point in opposing directions.
The core idea is to identify conflicts by calculating the cosine similarity between the gradients of different tasks. When a conflict is detected (cosine similarity is negative, indicating opposing directions), the gradient of the task is adjusted to ensure alignment by projecting it onto the normal plane of the conflicting gradient. If no conflict is found, the gradient remains unchanged.
By resolving these conflicts, PCGrad enables SeerNet-Multi to effectively predict multiple performance metrics simultaneously while maintaining accuracy.

\begin{table*}[!htb]
\belowrulesep=0pt
\aboverulesep=0pt
\newcolumntype{C}{>{\centering\arraybackslash}p{25pt}}
\renewcommand{\arraystretch}{1}
    \resizebox{2.0\columnwidth}{!}{
    \begin{tabular}{CCCCCCCCCCCCccccccc}
    
    \specialrule{1.5pt}{0pt}{0pt}
    \multicolumn{12}{c}{Feature} & \multicolumn{7}{c}{Accuracy (MAPE[\%]$\downarrow$)} \\
    \cmidrule(lr){1-12} \cmidrule(lr){13-19}
    
    \multicolumn{5}{c}{Node} & \multicolumn{2}{c}{Edge} & \multicolumn{5}{c}{Global}
    & \multicolumn{3}{c}{Training} & \multicolumn{3}{c}{Inference} 
    & \multirow{2}{*}{\makecell{\textit{Mean} \\ (6 metrics)}} \\
    \cmidrule(lr){1-5} \cmidrule(lr){6-7} \cmidrule(lr){8-12} \cmidrule(lr){13-15} \cmidrule(lr){16-18}
    
    Hp & CI & MAI & ArI* & \multicolumn{1}{c}{Prop*} & Sz & \multicolumn{1}{c}{Sp} & Gp* & CI* & \multicolumn{1}{c}{MAI*} &  ArI* & Bs & Util & Mem & \multicolumn{1}{c}{Time} & Util & Mem & \multicolumn{1}{c}{Time} \\
    \specialrule{1.0pt}{0pt}{0pt}

    \cm & \cu & \cu & \cu & \cu & \cu & \cu & \cu & \cu & \cu & \cu & \cu & 71.33 & 49.56 & 54.29 & 57.04 & 24.69 & 92.49 & \textcolor{colornode}{\textit{58.23}} \\
    \cm & \cm & \cu & \cu & \cu & \cu & \cu & \cu & \cu & \cu & \cu & \cu & 73.87 & 63.58 & 41.19 & 57.15 & 30.73 & 54.07 & \textcolor{colornode}{\textit{53.43}} \\
    \cm & \cm & \cm & \cu & \cu & \cu & \cu & \cu & \cu & \cu & \cu & \cu & 53.71 & 37.01 & 25.98 & 46.03 & 11.45 & 38.29 & \textcolor{colornode}{\textit{35.41}} \\
    \cm & \cm & \cm & \cm & \cu & \cu & \cu & \cu & \cu & \cu & \cu & \cu & 37.20 & 36.81 & 23.94 & 38.57 & 11.67 & 36.52 & \textcolor{colornode}{\textit{30.79}} \\
    \cm & \cm & \cm & \cm & \cm & \cu & \cu & \cu & \cu & \cu & \cu & \cu & 38.85 & 36.76 & 23.51 & 36.32 & 11.51 & 23.51 & \textcolor{colornode}{\textit{28.41}} \\
    \cmidrule(lr){1-19}
    
    \cm & \cm & \cm & \cm & \cm & \cm & \cu & \cu & \cu & \cu & \cu & \cu & 41.24 & 25.29 & 24.05 & 37.55 & 9.28 & 23.27 & \textcolor{coloredge}{\textit{26.78}} \\
    \cm & \cm & \cm & \cm & \cm & \cm & \cm & \cu & \cu & \cu & \cu & \cu & 36.02 & 27.11 & 22.31 & 33.27 & 9.93 & 23.77 & \textcolor{coloredge}{\textit{25.40}} \\
    \cmidrule(lr){1-19}
    
    \cm & \cm & \cm & \cm & \cm & \cm & \cm & \cm & \cu & \cu & \cu & \cu & 7.48 & 3.83 & 9.48 & 8.32 & 6.33 & 15.22 & \textcolor{colorglobal}{\textit{8.44}} \\
    \cm & \cm & \cm & \cm & \cm & \cm & \cm & \cm & \cm & \cu & \cu & \cu & 7.25 & 3.72 & 8.79 & 7.93 & 5.93 & 13.08 & \textcolor{colorglobal}{\textit{7.78}} \\
    \cm & \cm & \cm & \cm & \cm & \cm & \cm & \cm & \cm & \cm & \cu & \cu & 6.22 & 3.39 & 9.51 & 6.98 & 5.51 & 14.41 & \textcolor{colorglobal}{\textit{7.67}} \\
    \cm & \cm & \cm & \cm & \cm & \cm & \cm & \cm & \cm & \cm & \cm & \cu & 6.23 & 3.38 & 9.45 & 7.37 & 5.29 & 13.45 & \textcolor{colorglobal}{\textit{7.53}} \\
    \cm & \cm & \cm & \cm & \cm & \cm & \cm & \cm & \cm & \cm & \cm & \cm & 6.09 & 3.32 & 9.21 & 7.57 & 5.13 & 13.15 & \textcolor{colorglobal}{\textit{7.41}} \\
    
    \specialrule{1.5pt}{0pt}{0pt}

    \end{tabular}}
\caption{Ablation study of features. 
Hp, CI, MAI, ArI, Prop, Sz, Sp, GP, and Bs represent hyper-parameters, arithmetic intensity, computation info, memory access info, proportions, size, shape, graph profile, and batch size, respectively. "*" indicates our uniquely constructed features.}
\label{tab:ablation of features}
\end{table*}

\section{EVALUATION}
We evaluate PerfSeer by addressing the following research questions (RQs).

\textbf{RQ1:} How effective are the selected features and optimization components? 

\textbf{RQ2:} How does SeerNet compare with baseline models? 

\textbf{RQ3:} How effective is the multi-metric performance prediction model, SeerNet-Multi? 

\textbf{RQ4:} What is the application scope and overhead of PerfSeer? 

\subsection{Evaluation Setup}
\subsubsection{Training Settings.}\label{sec:train-setting}
The dataset is divided into 2:1:1 for training, validation, and testing. We use a batch size of 128 and an initial learning rate of 1e-3, halving it after five epochs without improvement, down to 1e-6. Training runs for up to 500 epochs, with Mean Squared Error (MSE) as the loss function and Adam as the optimizer.

\subsubsection{Evaluation Metrics.}\label{sec:metrics}
To ensure consistency across metrics with varying value ranges, we use percentage errors. The metrics include MAPE, Root Mean Square Percentage Error (RMSPE), and accuracy within a relative error of \(x\%\) (x\%Acc) \cite{Brp-nas}.

\subsection{Ablation Study (RQ1)}\label{sec:ablation}
The accuracy of the performance prediction depends on both representation and prediction. We conducted an ablation study to validate the features selection and construction (\emph{representation}) and the prediction model design (\emph{prediction}).
\subsubsection{Ablation Study of Features.} \label{sec:features ablation}
From the results shown in Table \ref{tab:ablation of features}, we observe the following:

\emph{Features Importance.} 
Node features are much more important than global features, which are in turn more important than edge features. 
The mean MAPE of SeerNet decreases from 58.23\% to 28.41\% as the node features are used increasingly, as they directly correspond to each operation node.
Adding the global features further improves accuracy, significantly reducing the mean MAPE from 25.40\% to 7.41\%, as they provide an overall representation of the model.
Edge features, though less important, improved MAPE slightly from 28.41\% to 25.40\% by providing additional memory access information, further enhancing prediction accuracy.

\emph{Feature Selection.} 
Each group of the node, edge, and global features we selected improved the performance prediction accuracy, demonstrating their relevance to model performance.
In contrast, node categories, which are commonly used in other predictors, reduced accuracy. We consider SeerNet extracts category-related information from other semantically rich features, so we excluded node categories from our feature set.

\emph{Feature Construction.} 
We constructed several unique and effective features. 
For node features, we included arithmetic intensity and computation/memory access proportions, reducing the mean MAPE from 35.41\% to 28.41\%. 
For global features, we introduced graph profiles, computation and memory access trend statistics, tensor size delivered per-edge, and arithmetic intensity, reducing the mean MAPE from 25.40\% to 7.53\%. 
These features, not used by any previous predictors, significantly enhance prediction accuracy.

\subsubsection{Ablation Study of Components.}\label{sec:ablation component}
From the results shown in Table \ref{tab:ablation of components}, we observe the following:

\emph{SynMM.}
Using SynMM to aggregate the node features reduces the mean MAPE from 7.41\% to 6.10\%. This demonstrates that SynMM combines max and mean aggregation to create a more comprehensive and robust representation, thereby enhancing prediction accuracy.

\emph{GNPB.}
With the introduction of GNPB, the mean MAPE further decreases from 6.10\% to 5.14\%. This result proves that GNPB enables complementary learning between the node and global features, enriching both perspectives and further improving prediction accuracy.

\begin{table}[!htb]
\belowrulesep=0pt
\aboverulesep=0pt
\renewcommand{\arraystretch}{1}{
    \resizebox{1.0\columnwidth}{!}
    {\begin{tabular}{ccccccccc}
    
    \specialrule{1.5pt}{0pt}{0pt}
    \multicolumn{2}{c}{Component} & \multicolumn{7}{c}{Accuracy(MAPE[\%]$\downarrow$)} \\
    \cmidrule(lr){1-2} \cmidrule(lr){3-9}

    \multirow{2}{*}{SynMM} & \multirow{2}{*}{GNPB} & \multicolumn{3}{c}{Training} & \multicolumn{3}{c}{Inference} & \multirow{2}{*}{\makecell{\textit{Mean} \\ (6 metrics)}} \\
    \cmidrule(lr){3-8}
    
    & & Util & Mem & \multicolumn{1}{c}{Time} & Util & Mem & \multicolumn{1}{c}{Time} \\
    \specialrule{1pt}{0pt}{0pt}

    \cu & \cu & 6.09 & 3.32 & 9.21 & 7.57 & 5.13 & 13.15 & \textit{7.41} \\
    \cm & \cu & 6.13 & 2.54 & 7.78 & 5.80 & 3.62 & 10.70 & \textit{6.10} \\
    \cm & \cm & \textbf{4.94} & \textbf{2.47} & \textbf{6.71} & \textbf{4.37} & \textbf{3.46} & \textbf{8.91} & \textit{\textbf{5.14}} \\
    \specialrule{1.5pt}{0pt}{0pt}
    
    \end{tabular}}}
\caption{Ablation study components.}
\label{tab:ablation of components}
\end{table}

\begin{table}[!htb]

\belowrulesep=0pt
\aboverulesep=0pt
\renewcommand{\arraystretch}{1}
    \resizebox{1.0\columnwidth}{!}{\begin{tabular}{lcccccccc}
    
    \specialrule{1.5pt}{0pt}{0pt}
    \multirow{3}{*}{Method} & \multicolumn{7}{c}{Accuracy (MAPE[\%]$\downarrow$)} & \multirow{3}{*}{\makecell{Params$\downarrow$ \\ {[}M{]}}} \\
    \cmidrule(lr){2-8}
    
    & \multicolumn{3}{c}{Training} & \multicolumn{3}{c}{Inference} & \multirow{2}{*}{\makecell{\textit{Mean} \\ (6 metrics)}}  \\
    \cmidrule(lr){2-4} \cmidrule(lr){5-7}
    
    & {Util} & {Mem} & {Time} & {Util} & {Mem} & {Time} & \multicolumn{1}{c}{} \\
    \specialrule{1.0pt}{0pt}{0pt}
    
    MLP-Node (MLP) & 21.27 & 5.41 & 12.37 & 35.21 & 6.38 & 23.77 & \textit{17.40} & 4.15 \\ 
    PMGNS (GraphSAGE) & 8.23 & 9.62 & 10.53 & 6.71 & 8.76 & 13.72 & \textit{9.60} & 3.45 \\ 
    Eagle-p (GCN) & 82.45 & 82.30 & 66.66 & 71.85 & 48.51 & 94.80 & \textit{74.43} & 1.11 \\ 
    Eagle-s (GCN) & 6.14 & 4.25 & 8.60 & 5.22 & 5.27 & 16.63 & \textit{7.69} & 1.10 \\ 
    \textbf{SeerNet (Our)} & \textbf{4.94} & \textbf{2.47} & \textbf{6.71} & \textbf{4.37} & \textbf{3.46} & \textbf{8.91} & \textbf{\textit{5.14}} & \textbf{1.02} \\ 

    \specialrule{1.5pt}{0pt}{0pt}
    
  \end{tabular}}
\caption{SeerNet comparison on our dataset}
\label{tab:comparison our}
\end{table}

\begin{table*}[!htb]
\belowrulesep=0pt
\aboverulesep=0pt
\newcolumntype{C}{>{\centering\arraybackslash}p{30pt}}
\renewcommand{\arraystretch}{1}
    \resizebox{2.0\columnwidth}{!}{
    \begin{tabular}{llCCCCCCCCCCCC}

    \specialrule{1.5pt}{0pt}{0pt}
    \multirow{3}{*}{Method} & \multirow{3}{*}{Model} & \multicolumn{3}{c}{Mobile CPU (CortexA76)} & \multicolumn{3}{c}{Mobile GPU (Adreno 640)} & \multicolumn{3}{c}{Intel VPU (MyriadX)} & \multicolumn{3}{c}{\textit{Mean (3 devices)}} \\
    
    & & RMSPE$\downarrow$ & Acc$\uparrow$ & Acc$\uparrow$ & RMSPE$\downarrow$ & Acc$\uparrow$ & Acc$\uparrow$ & RMSPE$\downarrow$ & Acc$\uparrow$ & Acc$\uparrow$ & RMSPE$\downarrow$ & Acc$\uparrow$ & Acc$\uparrow$ \\
    & & [\%] & [5\%] & [10\%] & [\%] & [5\%] & [10\%] & [\%] & [5\%] & [10\%] & [\%] & [5\%] &  [10\%] \\
    \specialrule{1.5pt}{0pt}{0pt}
    
    \multirow{13}{*}{\textit{nn-Meter}} & AlexNets & 3.90 & 81.0 & 98.6 & 5.32 & 72.0 & 94.0 & 10.74 & 23.4 & 60.9 & \textit{6.65} & \textit{58.8} & \textit{84.5} \\
    & DenseNets & 2.76 & 93.1 & 99.9 & 4.52 & 68.6 & 99.9 & 5.89 & 75.6 & 86.3 & \textit{4.39} & \textit{79.1} & \textit{95.4} \\
    & GoogleNets & 3.27 & 85.9 & 100.0 & 1.35 & 100.0 & 100.0 & 5.86 & 39.7 & 98.4 & \textit{3.49} & \textit{75.2} & \textit{99.5} \\
    & MnasNets & 5.54 & 50.9 & 99.2 & 1.86 & 100.0 & 100.0 & 4.34 & 77.3 & 97.7 & \textit{3.91} & \textit{76.1} & \textit{99.0} \\
    & MobileNetv1s & 4.98 & 63.8 & 97.8 & 2.56 & 96.9 & 100.0 & 5.90 & 54.2 & 93.3 & \textit{4.48} & \textit{71.6} & \textit{97.0} \\
    & MobileNetv2s & 4.84 & 67.6 & 97.7 & 3.93 & 80.0 & 99.0 & 4.26 & 78.3 & 97.6 & \textit{4.34} & \textit{75.3} & \textit{98.1} \\
    & MobileNetv3s & 4.34 & 73.8 & 99.0 & 4.02 & 84.4 & 100.0 & 5.72 & 47.6 & 98.5 & \textit{4.69} & \textit{68.6} & \textit{99.1} \\
    & NASBench201 & 3.51 & 82.4 & 99.9 & 3.80 & 75.9 & 100.0 & 18.20 & 19.3 & 40.6 & \textit{8.50} & \textit{59.2} & \textit{80.1} \\
    & ProxylessNas & 3.44 & 84.6 & 100.0 & 3.28 & 95.6 & 98.9 & 5.05 & 65.6 & 96.9 & \textit{3.92} & \textit{81.9} & \textit{98.6} \\
    & ResNets & 4.41 & 72.3 & 98.1 & 3.16 & 88.8 & 99.9 & 7.42 & 37.9 & 84.2 & \textit{5.00} & \textit{66.3} & \textit{94.1} \\
    & ShuffleNetv2s & 5.01 & 61.6 & 98.3 & - & - & - & 6.37 & 45.6 & 91.3 & \textit{5.69} & \textit{53.6} & \textit{94.8} \\
    & SqueezeNets & 3.59 & 84.5 & 99.9 & 3.85 & 81.9 & 97.9 & 7.08 & 66.1 & 88.5 & \textit{4.84} & \textit{77.5} & \textit{95.4} \\
    & VGGs & 4.84 & 66.1 & 98.2 & 2.97 & 91.8 & 99.8 & 22.25 & 27.1 & 50.6 & \textit{10.02} & \textit{61.7} & \textit{82.9} \\
    & \textit{Mean (13 models)} & \textit{4.19}& \textit{74.4} & \textit{99.0} & \textit{3.39} & \textit{86.3} & \textit{99.1} & \textit{8.39} & \textit{50.6} & \textit{83.4} & \underline{\textit{5.38}} & \underline{\textit{69.6}} & \underline{\textit{93.7}} \\
    \specialrule{1.0pt}{0pt}{0pt}
    
    \multirow{13}{*}{\textit{SeerNet (Our)}}
    & AlexNets & 3.91$\dagger$ & 83.3 & 97.9$\dagger$ & 3.34 & 89.6 & 97.7 & 3.48 & 84.6 & 99.5 & \textit{3.58} & \textit{85.9} & \textit{98.4} \\
    & DenseNets & 2.33 & 95.8 & 100.0 & 1.15 & 100.0 & 100.0 & 1.82 & 99.2 & 100.0 & \textit{1.77} & \textit{98.4} & \textit{100.0} \\
    & GoogleNets & 2.04 & 99.0 & 100.0 & 1.06 & 100.0 & 100.0 & 1.36 & 100.0 & 100.0 & \textit{1.49} & \textit{99.7} & \textit{100.0} \\
    & MnasNets & 2.49 & 97.9 & 100.0 & 1.70 & 100.0 & 100.0 & 3.23 & 88.8 & 99.0 & \textit{2.47} & \textit{95.6} & \textit{99.7} \\
    & MobileNetv1s & 2.53 & 96.4 & 100.0 & 1.47 & 100.0 & 100.0 & 3.71 & 84.9 & 98.2 & \textit{2.57} & \textit{93.8} & \textit{99.4} \\
    & MobileNetv2s & 3.18 & 87.8 & 100.0 & 2.66 & 93.2 & 100.0 & 4.13 & 78.4 & 98.7 & \textit{3.32} & \textit{86.5} & \textit{99.6} \\
    & MobileNetv3s & 2.80 & 93.0 & 100.0 & 2.23 & 96.9 & 100.0 & 2.06 & 98.2 & 100.0 & \textit{2.36} & \textit{96.0} & \textit{100.0} \\
    & NasBench201s & 2.77 & 94.8 & 100.0 & 2.32 & 97.7 & 100.0 & 3.74 & 82.3 & 98.2 & \textit{2.94} & \textit{91.6} & \textit{99.4} \\
    & ProxylessNas & 2.40 & 97.4 & 100.0 & 2.07 & 97.4 & 100.0 & 1.93 & 97.9 & 100.0 & \textit{2.13} & \textit{97.6} & \textit{100.0} \\
    & ResNets & 4.51$\dagger$ & 76.4 & 96.3$\dagger$ & 2.72 & 90.9 & 99.4$\dagger$ & 3.08 & 88.9 & 99.4 & \textit{3.44} & \textit{85.4} & \textit{98.4} \\
    & ShuffleNetv2s & 2.69 & 95.6 & 100.0 & - & - & - & 2.35 & 96.9 & 100.0 & \textit{2.52} & \textit{96.2} & \textit{100.0} \\
    & SqueezeNets & 3.33 & 87.0 & 100.0 & 2.36 & 96.9 & 100.0 & 4.44 & 75.8 & 96.6 & \textit{3.38} & \textit{86.6} & \textit{98.9} \\
    & VGGs & 6.03$\dagger$ & 61.9 & 93.2$\dagger$ & 2.44 & 94.9 & 99.4$\dagger$ & 13.52 & 38.4 & 62.2 & \textit{7.33} & \textit{65.1} & \textit{84.9} \\
    & \textit{Mean (13 models)} & \textit{3.15} & \textit{89.7} & \textit{99.0} & \textit{2.13} & \textit{96.5} & \textit{99.7} & \textit{3.76} & \textit{85.7} & \textit{96.3} & \underline{\textit{3.02}} & \underline{\textit{90.6}} & \underline{\textit{98.4}} \\
    \specialrule{1.5pt}{0pt}{0pt}
    
  \end{tabular}}
\caption{Comparison with nn-Meter. "$\dagger$" indicates SeerNet underperforms than nn-Meter, and the italicized and underlined entries represent the prediction results of execution time across 13 model types on 3 devices.}
\label{tab:comparison nnm}
\end{table*}
\subsection{Baseline Comparison (RQ2)}\label{sec:seernet compare}
\subsubsection{Baseline.} 
We compare SeerNet with the following methods:
(i) MLP-Node uses an MLP to predict model performance, concatenating features from all nodes in the graph and handling variable node numbers by padding or truncating them to a fixed size.
(ii) PMGNS \cite{DIPPM} (described in Section \ref{sec:related work}) utilizes a single prediction head for predicting one performance metric.
(iii) Eagle \cite{Brp-nas} (described in Section \ref{sec:related work}) is implemented for cell-based models, but for non-cell-based models, it uses features from PMGNS (Eagle-p) and our features from SeerPerf (Eagle-s).
(iv) nn-Meter \cite{Nn-meter} (described in Section \ref{sec:related work}) is a kernel-based predictor.

\subsubsection{Performance Comparison.}
\emph{Baseline Comparison on our Dataset.}
Table \ref{tab:comparison our} shows that SeerNet has the smallest parameter size (1.02M) and the highest accuracy, with a mean MAPE of 5.14\%.
Compared to SeerNet,
MLP-Node contains more than four times the parameters and achieves a MAPE over three times higher, while PMGNS contains more than three times the parameters and has a MAPE nearly twice as high.
Eagle-p performs poorly, with a MAPE of 74.43\%, due to the inability of PMGNS features to accurately represent the models in our dataset.
Eagle-s, which uses our proposed features, performs better but still lags behind SeerNet.
Both PMGNS and Eagle-s outperform MLP-Node, achieving higher accuracy with fewer parameters, highlighting the ability of GNNs to capture execution dependencies.
Eagle-s also performs better than Eagle-p, demonstrating the effectiveness of our proposed features. SeerNet outperforms the other models, offering the best representation and prediction.

\emph{Baseline Comparison on the Dataset of nn-Meter.} 
Table \ref{tab:comparison nnm} shows SeerNet outperforms nn-Meter with half the RMSPE, 21\% higher at 5\%Acc, and 5\% higher at 10\%Acc.
For models like VGG, ResNet, and AlexNet on the CPU, SeerNet is slightly less accurate than nn-Meter, likely due to the simple structures of these models and their predictable execution patterns on the CPU.
On VPU, nn-Meter achieves 50.6\% at 5\%Acc, while SeerNet reaches 85.7\%, 35\% higher. 
This is because the execution of VPU is more complex, and nn-Meter fails to design effective detection functions. 

\begin{table}[!htb]

\belowrulesep=0pt
\aboverulesep=0pt
\renewcommand{\arraystretch}{1}
    \resizebox{1.0\columnwidth}{!}{\begin{tabular}{lcccccccc}
    
    \specialrule{1.5pt}{0pt}{0pt}
    \multirow{3}{*}{Method} & \multicolumn{7}{c}{Accuracy (MAPE[\%]$\downarrow$)} & \multirow{3}{*}{\makecell{Params$\downarrow$ \\ {[}M{]}}} \\
    \cmidrule(lr){2-8}
    
    & \multicolumn{3}{c}{Training} & \multicolumn{3}{c}{Inference} & \multirow{2}{*}{\makecell{\textit{Mean} \\ (6 metrics)}}  \\
    \cmidrule(lr){2-4} \cmidrule(lr){5-7}

    & {Util} & {Mem} & {Time} & {Util} & {Mem} & {Time} & \multicolumn{1}{c}{} \\
    \specialrule{1.0pt}{0pt}{0pt}

    PMGNS-Multi & 38.7 & 10.6 & 21.7 & 83.4 & 48.7 & 99.8 & \textit{50.5} & 3.45 \\ 
    SeerNet (×3) & 4.94 & 2.47 & 6.71 & 4.37 & 3.46 & 8.91 & \textit{5.14} & 3.05 \\ 
    SeerNet-Multi (w/o PCGrad) & 17.60 & 3.00 & 22.10 & 18.90 & 3.70 & 29.90 & \textit{15.85} & 1.15 \\ 
    \textbf{SeerNet-Multi (w/ PCGrad)} & \textbf{6.90} & \textbf{3.30} & \textbf{9.10} & \textbf{8.70} & \textbf{3.60} & \textbf{15.20} & \textbf{\textit{7.75}} & \textbf{1.15} \\ 
    \specialrule{1.5pt}{0pt}{0pt}
    
    \end{tabular}}
\caption{Evaluation result of SeerNet-Multi.}
\label{tab:seernet multi}
\end{table}

\begin{table}[!htb]

\belowrulesep=0pt
\aboverulesep=0pt
\newcolumntype{C}{>{\centering\arraybackslash}p{30pt}}
\renewcommand{\arraystretch}{1}
    \resizebox{1.0\columnwidth}{!}{\begin{tabular}{rccccccc}
    
    \specialrule{1.5pt}{0pt}{0pt}
    \multirow{3}{*}{\makecell{Dataset \\ scale}} & \multicolumn{7}{c}{Accuracy (MAPE[\%]$\downarrow$)} \\
    \cmidrule(lr){2-8}
    
    & \multicolumn{3}{c}{Training} & \multicolumn{3}{c}{Inference} & \multirow{2}{*}{\makecell{Mean \\ (6 metrics)}} \\
    \cmidrule(lr){2-4} \cmidrule(lr){5-7}
    
    & {Util} & {Mem} & {Time} & {Util} & {Mem} & {Time} \\
    \specialrule{1.0pt}{0pt}{0pt}

    \textbf{overall} & \textbf{4.94} & \textbf{2.47} & \textbf{6.71} & \textbf{4.37} & \textbf{3.46} & \textbf{8.91} & \textbf{\textit{5.14}} \\ 
    20000 & 5.49 & 2.49 & 8.58 & 5.04 & 3.83 & 13.65 & \textit{6.51} \\ 
    10000 & 6.01 & 2.91 & 10.05 & 5.31 & 5.37 & 14.49 & \textit{7.36} \\ 
    5000 & 7.61 & 4.24 & 10.78 & 6.32 & 4.79 & 17.89 & \textit{8.61} \\ 
    2000 & 10.24 & 6.39 & 12.33 & 8.26 & 7.49 & 19.25 & \textit{10.66} \\ 
    1000 & 18.70 & 5.53 & 15.68 & 14.86 & 6.94 & 28.19 & \textit{14.98} \\ 
    
  \specialrule{1.5pt}{0pt}{0pt}
  
  \end{tabular}}
\caption{Data dependency of SeerNet.}
\label{tab:data dependency}
\end{table}

\subsection{Effectiveness of SeerNet-Multi (RQ3)}\label{sec:eva seernet multi}
Table \ref{tab:comparison nnm} shows that PMGNS-Multi (PMGNS with multiple prediction heads) performed poorly with a MAPE of 50.5\%, while SeerNet-Multi (without PCGrad) had a MAPE of 15.85\%, indicating that predicting multiple metrics simultaneously reduces accuracy.
However, with PCGrad, SeerNet-Multi halved its MAPE from 15.85\% to 7.75\% without increasing parameter overhead. This demonstrates that PCGrad effectively mitigates conflicting gradient directions across tasks, enabling SeerNet-Multi to predict multiple metrics efficiently with minimal accuracy loss.
Furthermore, SeerNet-Multi has about one-third the parameters of SeerNet, with only a 2.61\% increase in MAPE, balancing parameter efficiency and prediction accuracy. This makes SeerNet-Multi ideal for scenarios requiring rapid predictions with limited resources and lower accuracy demands.

\subsection{Further Discussion (RQ4)}\label{sec:extension experiments}
\subsubsection{Application Scope.}
\emph{Multi-Model, Multi-Metric Support.}
SeerPerf provides accurate predictions for execution time, memory usage, and SM utilization during both training and inference across various architectures, including GoogLeNet, VGG, ResNe(X)t, MobileNet, and DenseNet.

\emph{Multi-Device Support.}
PerfSeer provides accurate predictions across various devices, including mobile CPUs, mobile GPUs, desktop GPUs, and Intel VPUs, as shown in Table \ref{tab:comparison nnm}. In contrast, nn-Meter exhibits poor prediction accuracy on Intel VPUs.

\emph{Multi-Platform Support.}
The representation of PerfSeer is based on ONNX, so SeerPerf supports any DL framework convertible to ONNX.
\emph{Overall}, our performance predictor, PerfSeer, has a wide application scope, making it suitable for most common applications.

\subsubsection{Overhead.}
\emph{Data dependency and deployment overhead.}
To evaluate the data dependency of SeerNet, we analyze the relationship between dataset scale and prediction accuracy, keeping the test set size fixed. 
Results (Table \ref{tab:data dependency}) show that accuracy decreases as the dataset scale shrinks. Nevertheless, SeerNet achieves a mean MAPE of 14.98 with only 1,000 samples, demonstrating low data dependency.
The deployment overhead includes 16.67 GPU hours for data collection and 0.05 GPU hours for training per 1,000 samples, resulting in low deployment overhead.

\emph{Usuage overhead.}
We evaluated the overhead of SeerPerf on an Intel i7-11700 CPU, which includes representation and prediction. 
The average representation latency is 248 ms, with prediction latencies of 2.0 ms for SeerNet and 2.1 ms for SeerNet-Multi. The total overhead of approximately 250 ms is acceptable for most applications. \emph{Overall}, SeerPerf demonstrates low overhead in both deployment and usage.


\section{CONCLUSION}
We propose PerfSeer, a novel predictor that efficiently and accurately predicts key performance metrics for models during both training and inference. 
To accurately predict the performance, PerfSeer abstracts a model into a graph with the node, edge, and global features and uses SeerNet, which leverages these features and incorporates optimizations like SynMM and GNPB.
To effectively predict multiple performance metrics simultaneously while maintaining accuracy, PerfSeer implements SeerNet-Multi with PCGrad. 
The evaluation results show that PerfSeer has low overhead and high prediction accuracy, making it suitable for broad applications.


\bibliographystyle{named}
\bibliography{reference/nas, reference/scheduler, reference/predictor, reference/mlabout, reference/gnn, reference/mtl}

\begin{thebibliography}{}

\bibitem[\protect\citeauthoryear{Banbury \bgroup \em et al.\egroup }{2021}]{micronets-nas}
Colby Banbury, Chuteng Zhou, Igor Fedorov, Ramon~Matas Navarro, Urmish Thakker, Dibakar Gope, Vijay~Janapa Reddi, Matthew Mattina, and Paul~N. Whatmough.
\newblock Micronets: Neural network architectures for deploying tinyml applications on commodity microcontrollers, 2021.

\bibitem[\protect\citeauthoryear{Battaglia \bgroup \em et al.\egroup }{2018}]{gn}
Peter~W Battaglia, Jessica~B Hamrick, Victor Bapst, Alvaro Sanchez-Gonzalez, Vinicius Zambaldi, Mateusz Malinowski, Andrea Tacchetti, David Raposo, Adam Santoro, Ryan Faulkner, et~al.
\newblock Relational inductive biases, deep learning, and graph networks.
\newblock {\em arXiv preprint arXiv:1806.01261}, 2018.

\bibitem[\protect\citeauthoryear{Chai \bgroup \em et al.\egroup }{2023}]{PerfSAGE}
Yuji Chai, Devashree Tripathy, Chuteng Zhou, Dibakar Gope, Igor Fedorov, Ramon Matas, David Brooks, Gu-Yeon Wei, and Paul Whatmough.
\newblock Perfsage: Generalized inference performance predictor for arbitrary deep learning models on edge devices.
\newblock {\em arXiv preprint arXiv:2301.10999}, 2023.

\bibitem[\protect\citeauthoryear{Chen and Guestrin}{2016}]{Xgboost}
Tianqi Chen and Carlos Guestrin.
\newblock Xgboost: A scalable tree boosting system.
\newblock In {\em Proceedings of the 22nd acm sigkdd international conference on knowledge discovery and data mining}, pages 785--794, 2016.

\bibitem[\protect\citeauthoryear{Dong and Yang}{2020}]{nas-bench}
Xuanyi Dong and Yi~Yang.
\newblock Nas-bench-201: Extending the scope of reproducible neural architecture search.
\newblock In {\em International Conference on Learning Representations}, 2020.

\bibitem[\protect\citeauthoryear{Dudziak \bgroup \em et al.\egroup }{2020}]{Brp-nas}
Lukasz Dudziak, Thomas Chau, Mohamed Abdelfattah, Royson Lee, Hyeji Kim, and Nicholas Lane.
\newblock Brp-nas: Prediction-based nas using gcns.
\newblock {\em Advances in Neural Information Processing Systems}, 33:10480--10490, 2020.

\bibitem[\protect\citeauthoryear{Eriksson \bgroup \em et al.\egroup }{2021}]{latencyaware-nas}
David Eriksson, Pierce I-Jen Chuang, Samuel Daulton, Peng Xia, Akshat Shrivastava, Arun Babu, Shicong Zhao, Ahmed Aly, Ganesh Venkatesh, and Maximilian Balandat.
\newblock Latency-aware neural architecture search with multi-objective bayesian optimization, 2021.

\bibitem[\protect\citeauthoryear{Fedorov \bgroup \em et al.\egroup }{2019}]{sas-nas}
Igor Fedorov, Ryan~P. Adams, Matthew Mattina, and Paul~N. Whatmough.
\newblock Sparse: Sparse architecture search for cnns on resource-constrained microcontrollers, 2019.

\bibitem[\protect\citeauthoryear{Gao \bgroup \em et al.\egroup }{2023}]{dnnperf}
Yanjie Gao, Xianyu Gu, Hongyu Zhang, Haoxiang Lin, and Mao Yang.
\newblock Runtime performance prediction for deep learning models with graph neural network.
\newblock In {\em 2023 IEEE/ACM 45th International Conference on Software Engineering: Software Engineering in Practice (ICSE-SEIP)}, pages 368--380. IEEE, 2023.

\bibitem[\protect\citeauthoryear{Gilmer \bgroup \em et al.\egroup }{2017}]{virtualnode}
Justin Gilmer, Samuel~S. Schoenholz, Patrick~F. Riley, Oriol Vinyals, and George~E. Dahl.
\newblock Neural message passing for quantum chemistry.
\newblock In {\em Proceedings of the 34th International Conference on Machine Learning - Volume 70}, ICML'17, page 1263–1272. JMLR.org, 2017.

\bibitem[\protect\citeauthoryear{Gu \bgroup \em et al.\egroup }{2021}]{Liquid}
Rong Gu, Yuquan Chen, Shuai Liu, Haipeng Dai, Guihai Chen, Kai Zhang, Yang Che, and Yihua Huang.
\newblock Liquid: Intelligent resource estimation and network-efficient scheduling for deep learning jobs on distributed gpu clusters.
\newblock {\em IEEE Transactions on Parallel and Distributed Systems}, 33(11):2808--2820, 2021.

\bibitem[\protect\citeauthoryear{Hamilton \bgroup \em et al.\egroup }{2017}]{GraphSage}
Will Hamilton, Zhitao Ying, and Jure Leskovec.
\newblock Inductive representation learning on large graphs.
\newblock {\em Advances in neural information processing systems}, 30, 2017.

\bibitem[\protect\citeauthoryear{Haykin}{1998}]{mlp}
Simon Haykin.
\newblock {\em Neural networks: a comprehensive foundation}.
\newblock Prentice Hall PTR, 1998.

\bibitem[\protect\citeauthoryear{He \bgroup \em et al.\egroup }{2016}]{resnet}
Kaiming He, Xiangyu Zhang, Shaoqing Ren, and Jian Sun.
\newblock Deep residual learning for image recognition.
\newblock In {\em Proceedings of the IEEE conference on computer vision and pattern recognition}, pages 770--778, 2016.

\bibitem[\protect\citeauthoryear{Hopfield}{1982}]{RNN}
John~J Hopfield.
\newblock Neural networks and physical systems with emergent collective computational abilities.
\newblock {\em Proceedings of the national academy of sciences}, 79(8):2554--2558, 1982.

\bibitem[\protect\citeauthoryear{Howard \bgroup \em et al.\egroup }{2017}]{Mobilenets}
Andrew~G Howard, Menglong Zhu, Bo~Chen, Dmitry Kalenichenko, Weijun Wang, Tobias Weyand, Marco Andreetto, and Hartwig Adam.
\newblock Mobilenets: Efficient convolutional neural networks for mobile vision applications.
\newblock {\em arXiv preprint arXiv:1704.04861}, 2017.

\bibitem[\protect\citeauthoryear{Huang \bgroup \em et al.\egroup }{2017}]{densenet}
Gao Huang, Zhuang Liu, Laurens Van Der~Maaten, and Kilian~Q Weinberger.
\newblock Densely connected convolutional networks.
\newblock In {\em Proceedings of the IEEE conference on computer vision and pattern recognition}, pages 4700--4708, 2017.

\bibitem[\protect\citeauthoryear{Justus \bgroup \em et al.\egroup }{2018}]{justus}
Daniel Justus, John Brennan, Stephen Bonner, and Andrew~Stephen McGough.
\newblock Predicting the computational cost of deep learning models.
\newblock In {\em 2018 IEEE international conference on big data (Big Data)}, pages 3873--3882. IEEE, 2018.

\bibitem[\protect\citeauthoryear{Kipf and Welling}{2017}]{GCN}
Thomas~N. Kipf and Max Welling.
\newblock Semi-supervised classification with graph convolutional networks, 2017.

\bibitem[\protect\citeauthoryear{Krizhevsky \bgroup \em et al.\egroup }{2012}]{cnn}
Alex Krizhevsky, Ilya Sutskever, and Geoffrey~E Hinton.
\newblock Imagenet classification with deep convolutional neural networks.
\newblock {\em Advances in neural information processing systems}, 25, 2012.

\bibitem[\protect\citeauthoryear{Liaw \bgroup \em et al.\egroup }{2002}]{randomforest}
Andy Liaw, Matthew Wiener, et~al.
\newblock Classification and regression by randomforest.
\newblock {\em R news}, 2(3):18--22, 2002.

\bibitem[\protect\citeauthoryear{Panner~Selvam and Brorsson}{2023}]{DIPPM}
Karthick Panner~Selvam and Mats Brorsson.
\newblock Dippm: A deep learning inference performance predictive model using graph neural networks.
\newblock In {\em European Conference on Parallel Processing}, pages 3--16, 2023.

\bibitem[\protect\citeauthoryear{Simonyan and Zisserman}{2015}]{vggnet}
Karen Simonyan and Andrew Zisserman.
\newblock Very deep convolutional networks for large-scale image recognition.
\newblock In Yoshua Bengio and Yann LeCun, editors, {\em 3rd International Conference on Learning Representations, {ICLR} 2015, San Diego, CA, USA, May 7-9, 2015, Conference Track Proceedings}, 2015.

\bibitem[\protect\citeauthoryear{Szegedy \bgroup \em et al.\egroup }{2015}]{googlenet}
Christian Szegedy, Wei Liu, Yangqing Jia, Pierre Sermanet, Scott Reed, Dragomir Anguelov, Dumitru Erhan, Vincent Vanhoucke, and Andrew Rabinovich.
\newblock Going deeper with convolutions.
\newblock In {\em Proceedings of the IEEE conference on computer vision and pattern recognition}, pages 1--9, 2015.

\bibitem[\protect\citeauthoryear{Velickovic \bgroup \em et al.\egroup }{2017}]{GAT}
Petar Velickovic, Guillem Cucurull, Arantxa Casanova, Adriana Romero, Pietro Lio, Yoshua Bengio, et~al.
\newblock Graph attention networks.
\newblock {\em stat}, 1050(20):10--48550, 2017.

\bibitem[\protect\citeauthoryear{Xie \bgroup \em et al.\egroup }{2017}]{resnext}
Saining Xie, Ross Girshick, Piotr Doll{\'a}r, Zhuowen Tu, and Kaiming He.
\newblock Aggregated residual transformations for deep neural networks.
\newblock In {\em Proceedings of the IEEE conference on computer vision and pattern recognition}, pages 1492--1500, 2017.

\bibitem[\protect\citeauthoryear{Yeung \bgroup \em et al.\egroup }{2021}]{Horus}
Gingfung Yeung, Damian Borowiec, Renyu Yang, Adrian Friday, Richard Harper, and Peter Garraghan.
\newblock Horus: Interference-aware and prediction-based scheduling in deep learning systems.
\newblock {\em IEEE Transactions on Parallel and Distributed Systems}, 33(1):88--100, 2021.

\bibitem[\protect\citeauthoryear{Yu \bgroup \em et al.\egroup }{2020}]{pcgrad}
Tianhe Yu, Saurabh Kumar, Abhishek Gupta, Sergey Levine, Karol Hausman, and Chelsea Finn.
\newblock Gradient surgery for multi-task learning.
\newblock In {\em Proceedings of the 34th International Conference on Neural Information Processing Systems}, NIPS '20, page~13, Red Hook, NY, USA, 2020. Curran Associates Inc.

\bibitem[\protect\citeauthoryear{Zhang \bgroup \em et al.\egroup }{2021}]{Nn-meter}
Li~Lyna Zhang, Shihao Han, Jianyu Wei, Ningxin Zheng, Ting Cao, Yuqing Yang, and Yunxin Liu.
\newblock Nn-meter: Towards accurate latency prediction of deep-learning model inference on diverse edge devices.
\newblock In {\em Proceedings of the 19th Annual International Conference on Mobile Systems, Applications, and Services}, pages 81--93, 2021.

\end{thebibliography}

\end{document}